\documentclass[pra,twocolumn,amsmath,amssymb,superscriptaddress]{revtex4}
\usepackage{graphicx}

\newcommand{\ket}[1]{\ensuremath{\left |#1 \right \rangle}}

\renewcommand{\bf}[0]{}

\begin{document}


\date{\today}
\title{Magnetic trapping of silver and copper, and \\ anomalous spin relaxation in the Ag-He system
}

\author{Nathan Brahms}
\affiliation{Department of Physics, Harvard University, Cambridge, Massachusetts  02138}
\author{Bonna Newman}
\author{Cort Johnson}
\author{Tom Greytak}
\author{Daniel Kleppner}
\affiliation{Department of Physics, Massachusetts Institute of Technology, Cambridge, Massachusetts  02139}
\author{John Doyle}
\affiliation{Department of Physics, Harvard University, Cambridge, Massachusetts  02138}


\begin{abstract}
We have trapped large numbers of copper (Cu) and silver (Ag) atoms using buffer gas cooling.  Up to $3\times 10^{12}$ Cu atoms and $4\times 10^{13}$ Ag atoms are trapped.  Lifetimes are as long as 5 s, limited by collisions with the buffer gas.  Ratios of elastic to inelastic collision rates with He are $\gtrsim 10^6$, suggesting Cu and Ag are favorable for use in ultracold applications.  The temperature dependence of the Ag-$^3$He collision rate varies as $T^{5.8}$.
We find that this temperature dependence is inconsistent with the behavior predicted for relaxation arising from the spin-rotation interaction, and conclude that the Ag-3He system displays anomalous collisional behavior in the multiple-partial wave regime.
Lifetimes of laser ablated gold (Au) in $^3$He buffer gas are too short to permit trapping.
\end{abstract}



\maketitle



As interest in the physics of ultracold atoms and molecules continues to grow, there is a natural desire to expand the family of species that can be cooled and trapped.
New systems can exhibit novel collisional behavior at cold temperatures. {\bf Experiments can lead to} increased theoretical understanding of interatomic interactions, {\bf and have demonstrated, for example, suppressed collisional anisotropy in the rare earth elements} \cite{romanRareEarths,romanTransitionMetals2,doyleRareEarths,doyleTransitionMetals}.  
Like the alkali atoms, the noble metals are composed of closed core electron shells and a single valence $s$ electron, allowing the atoms to be magnetically trapped in their ground state.  This configuration is highly isotropic, suggesting that collisions between the atoms are {\bf almost always} elastic, and that the atoms can be cooled efficiently to ultracold temperatures using evaporative cooling.
At the same time, these atoms possess narrow two-photon transitions.  Silver has garnered particular interest: it has {\bf a} two-photon transition with a linewidth of \mbox{0.8 Hz}, which is interrogated at \mbox{661 nm}.  This line has attracted attention as a proposed frequency standard \cite{benderSilverClock}, with groups working on producing cold silver samples \cite{silverMot} and measuring the clock transition \cite{badrAgClockTransition}.

The large laser powers at UV frequencies needed to laser cool the noble metals (240 to \mbox{330 nm}) have posed a serious challenge to experiments thus far, limiting the number of silver atoms trapped in a MOT to $3\times 10^6$ \cite{silverMot}.  In this letter we describe the trapping of large numbers of noble metals {\bf and measurements of noble metal--He collisions}.
Using buffer gas cooling we trap large, dense samples of the noble metals copper (Cu) and silver (Ag) at temperatures as low as \mbox{300 mK} and with lifetimes as long as \mbox{5 s}, limited by collisions with the $^3$He buffer gas.
We study these collisions, discovering an anomalous temperature dependence for Ag-$^3$He inelastic collisions.  {\bf We compare this dependence to a standard theoretical prediction based on the spin-rotation interaction, showing a marked discrepancy between experiment and theory.}

\begin{figure}
\centering
\includegraphics[bb=0pt 0pt 200pt 246pt]{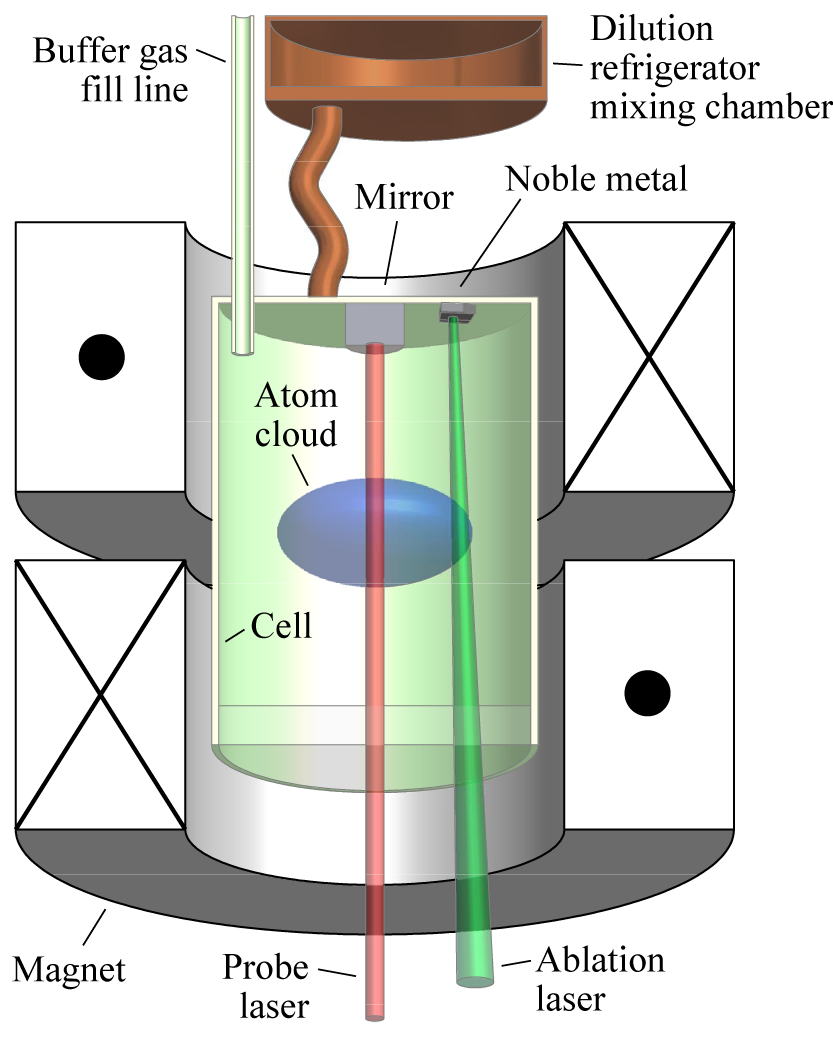}
\caption{Apparatus schematic.  The circles and crosses on the magnet coils indicate the current direction.}
\label{apparatusSchematicFigure}
\end{figure}

In our experiment, atoms are produced by laser ablation of the metallic state.  The ablation occurs within a cold cell (see Fig. \ref{apparatusSchematicFigure}) that is kept between \mbox{150 mK} and \mbox{1 K} by a dilution refrigerator.  The cell resides within the bore of a superconducting anti-Helmholtz magnet.  This creates a magnetic trap within the cell, with trap depth \mbox{4.0 T} at the cell walls.  The loss due to Majorana spin flips intrinsic to the trap is negligible at the temperatures and fields used in this experiment.  The cell is filled with gaseous $^3$He, chosen because it has a significant vapor pressure in this temperature range.  The ablated atoms thermalize to the temperature of the buffer gas $T_{\rm BG}$ within ~100 collisions.  Because this temperature is lower than the energy of interaction $E_{\rm trap}$ between the atoms and the magnetic field at the cell wall, atoms that are in a low-field seeking Zeeman state are trapped.  The lifetime of atoms in the trap is determined by the density of $^3$He in the cell.

At low He densities, elastic collisions between atoms and He can promote the atoms to untrapped energies.  These atoms can hit the cell walls, where they stick and are lost from the trap.  Loss from this mechanism is proportional to He density.  At larger He densities, the mean free path $\lambda$ for elastic collisions is smaller than the cell size $D_{\text{cell}}$.  At these densities, atoms promoted to untrapped energies can experience additional elastic collisions before they can travel to the cell wall.  Because $k_B T_{\rm BG} \ll E_{\rm trap}$, these additional collisions will typically return the atoms to trapped energies.  In this regime, the atomic loss rate decreases with increasing He density.  Finally, for even greater He densities, inelastic collisions between atoms and He can cause trapped low-field seeking atoms to change state to a high-field seeking Zeeman state.  These high-field seeking states are then ejected from the trap.  Our experiment is operated in the latter two density regimes: the multiple elastic collision regime and the inelastic collision regime.

In the absence of a magnetic trapping field, atoms diffuse through the He to the cell walls, where they stick and are lost from the cloud.  For atoms in the lowest-order diffusion mode, the $1/e$ lifetime of atoms in a cell of length $L$ and radius $R$ is \cite{hastedCollisions}
\begin{gather}
    \tau_0(n) = \frac {\bar\sigma_D n}{ g \bar v_\mu} \label{eqn:tau0}, \\
    g = \frac{3\pi}{32} \left (\frac{\pi^2}{L^2} + \frac{j_{01}^2}{R^2} \right ),
\end{gather}
where $\bar\sigma_D$ is the temperature-averaged transport cross-section for the atoms in $^3$He, $n$ is the density of $^3$He in the cell, and $\bar v_\mu$ is the average relative atom-$^3$He speed (close to the thermal velocity of $^3$He for heavy atoms).  $j_{01} = 2.40483\ldots$ is the first root of the Bessel function $J_0(z)$ \cite{besselZeros}.  Since the lowest-order diffusion mode has the longest lifetime, one can guarantee that most atoms are in the lowest-order diffusion mode simply by waiting some small number of lowest-order lifetimes. In practice, the decay of the atom number appears to be a single exponential for $t>\tau_0$.

{\bf To calculate the trap lifetime in the $D_{\text{cell}} > \lambda$ regime of He density,
we introduce the familiar dimensionless parameter $\eta = \mu B_{\rm trap}/{k_B T}$}.  $\mu$ is the dipole moment of the trapped atom, $B_{\text{trap}}$ is the minimum magnetic field at the cell wall, and $E_{\text{trap}} = \mu B_{\text{trap}}$.  We ran Monte-Carlo simulations to show that, for \mbox{$4 \le \eta \le 10$}, which includes the range studied in this experiment, atom lifetime in this {\bf density} regime can be related to $\tau_0(n)$:
\begin{equation}
    \tau_D(n,\eta) = \tau_0(n) e^{0.24 \eta+0.03 \eta^2}.
\label{eqn:tauD}
\end{equation}

The trapped atom lifetime due to inelastic spin relaxing collisions in the high He density regime is  
\begin{equation}
    \tau_R(n) = \frac 1 {n \bar\sigma_R \bar v_\mu},
\label{eqn:tauR}
\end{equation}
where $\bar\sigma_R$ is the temperature-averaged cross-section for atom-$^3$He Zeeman relaxing collisions.

The total lifetime of atoms in the trap is the reciprocal sum of these two lifetimes 
\begin{equation}
	\tau_{\text{trap}}(n) = \frac{1}{1/\tau_D(n,\eta) + 1/\tau_R(n)}.
\label{eqn:tauTUnsub}
\end{equation}

To study the collisional properties of the noble metals with $^3$He, we would like to use these lifetimes to determine $\bar\sigma_D (T)$ and $\bar\sigma_R (T)$.  Unfortunately we cannot precisely measure the density $n$ of He in the cell, which is necessary to determining either of these cross sections.  We can however, measure the ratio of the two cross sections, $\gamma(T) = \bar\sigma_D(T) / \bar\sigma_R(T)$, without knowing $n$.  $\gamma$ is a direct measure of the elasticity of collisions.  This number is useful both for accurate theoretical comparisons and for predicting the efficiency of evaporative cooling.

To eliminate $n$, we substitute (\ref{eqn:tau0}) and (\ref{eqn:tauD}) into (\ref{eqn:tauTUnsub}) , giving
\begin{equation}
    \tau_{\text{trap}}(\tau_0) = \frac{\tau_0}{e^{-0.24 \eta - 0.03 \eta^2} + ( g \bar v_{\mu}^2 / \gamma ) \tau_0^2 }.
\end{equation}
To measure $\gamma$, we load buffer gas into our cell to achieve a desired $\tau_0$.  We then turn on the magnetic trap to measure $\tau_{\text{trap}}$.  The experiment is repeated for a range of $\tau_0$, and $\tau_{\text{trap}}$ is fit to
\begin{equation}
    \tau_{\text{trap}}(\tau_0) = \frac {\tau_0}{A + B \tau_0^2},
    \label{eqn:tauTotal}
\end{equation}
where $A$ and $B$ are the fit parameters.  The cross-section ratio is determined from $B$, the geometry factor $g$, and the thermal velocity of the colliding system:
\begin{equation}
    \gamma = \frac{g \bar v_\mu^2}{B}.
\end{equation}
Note that the $\eta$ dependence calculated in (\ref{eqn:tauD}) does not enter into the expression for $\gamma$.

\begin{figure}
\centering
\includegraphics {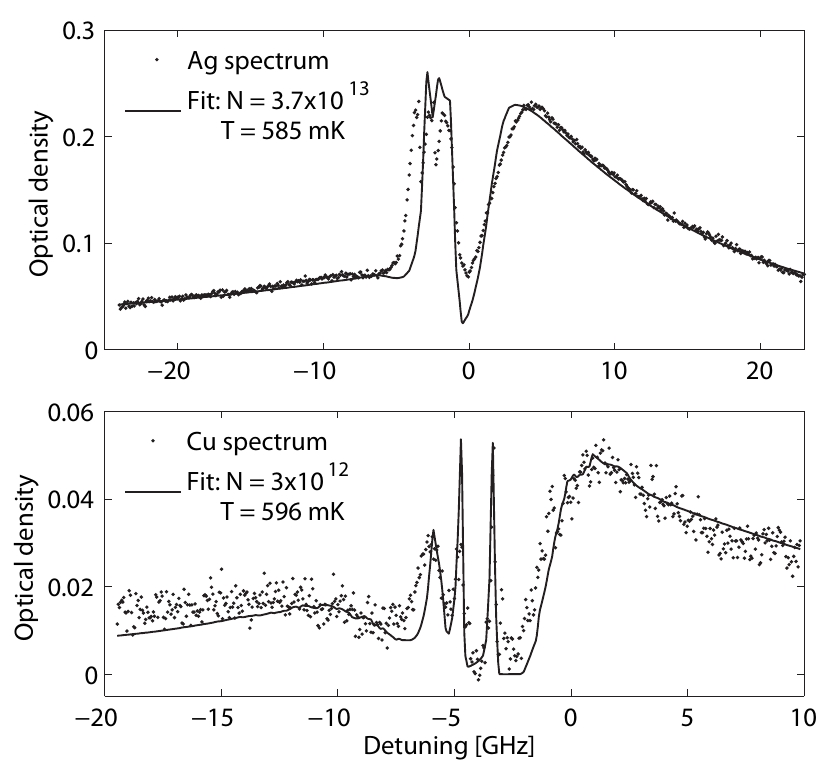}
\caption{Measured spectra of Cu and Ag in our 4 T anti-Helmholtz trap,  with best-fit simulations.  The fit parameters are number, temperature, probe beam profile and hyperfine sublevel populations.
\label{silverTrapFigure}}
\end{figure}

We probe our trapped atoms with absorption spectroscopy on the primary atomic transitions, \mbox{$^2S_{1/2}\rightarrow ^2J_{3/2}$}.  Typical spectra of trapped $\ket{m_J=\frac 12}$ Ag and Cu in our anti-Helmholtz trap are shown in Fig. \ref{silverTrapFigure}.
The spectra are simulated from experimental parameters including atom number, temperature, field profile, and probe beam characteristics \cite{spectrumSimulation,weinsteinThesis}.  By fitting the simulation to the measured data, we can extract properties of the trapped atom cloud.  We trapped {\bf up to} $4\times 10^{13}$ Ag atoms and $3\times 10^{12}$ Cu atoms {\bf at temperatures between 320 and 600 mK, and densities up to $1.3\times 10^{13}$ cc$^{-1}$}.  We attempted to trap Au, but encountered an unexplained loss process that limits the Au lifetime to \mbox{~70 ms}, regardless of the magnetic confinement.

\begin{figure}
\centering
\includegraphics {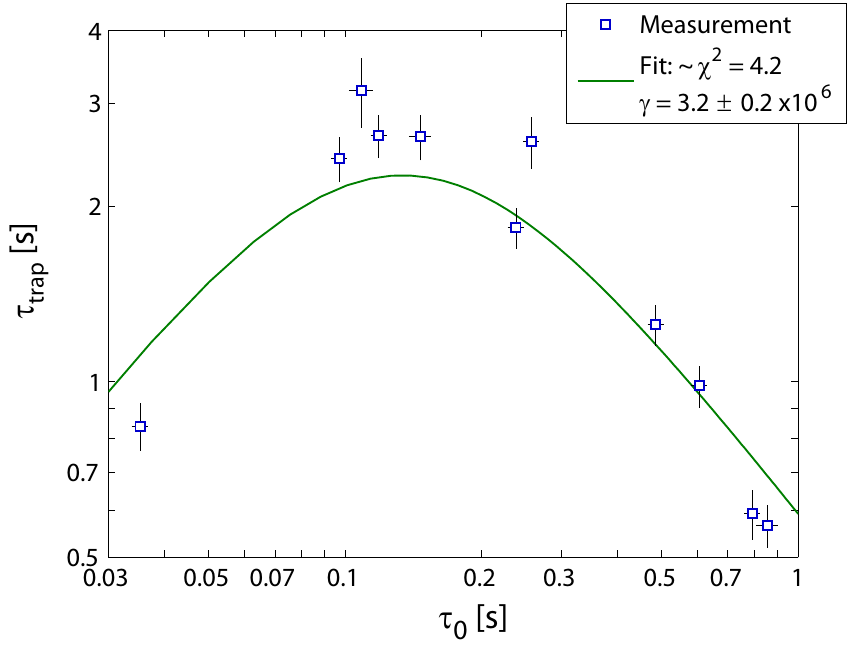}
\caption{Ag trap-on lifetime vs. trap-off lifetime at 420 mK.  The fit is to (\ref{eqn:tauTotal}).} \label{fig:lifetimeExample}
\end{figure}

Atom lifetime is determined by scanning the probe laser rapidly (between 5 and \mbox{100 Hz}) relative to the atom loss rate.  At a constant temperature, the spectrum is proportional to the atom number.  The atomic spectrum is integrated over each scan, and the integral vs. time is fit to an exponential decay.  Low-frequency noise is reduced by subtracting the mean of an off resonant portion of the spectrum from the integral.
Fig. \ref{fig:lifetimeExample} shows an example of the trap-on lifetime vs. the trap-off lifetime for Ag, along with a fit to (\ref{eqn:tauTotal}).  At low buffer gas densities, loss from elastic collisions is dominant, while at high buffer gas densities, spin relaxation is dominant.  At low He densities ($\tau_0 < 150$ ms), at \mbox{420 mK}, the trap enhances atom lifetime by a factor of 20.  By tuning the buffer gas density, we were able to trap Ag for up to \mbox{2.3 s} and Cu for up to \mbox{5 s}.  At \mbox{420 mK}, we measured $\gamma$ for Ag-$^3$He to be $3.2\pm 0.2\times 10^6$.  For the collision measurements we measure cell temperature using a solid-state thermometer.  This agrees with the atom temperature obtained from spectra to within \mbox{30 mK} over the time ranges for which we fit lifetimes.  The quoted uncertainty in $\gamma$ includes a 5\% systematic uncertainty associated with this temperature measurement.  At \mbox{310 mK}, the measured $\gamma$ for Cu-$^3$He was $8.2\pm 0.4\times 10^6$.

\begin{figure}
\centering
\includegraphics {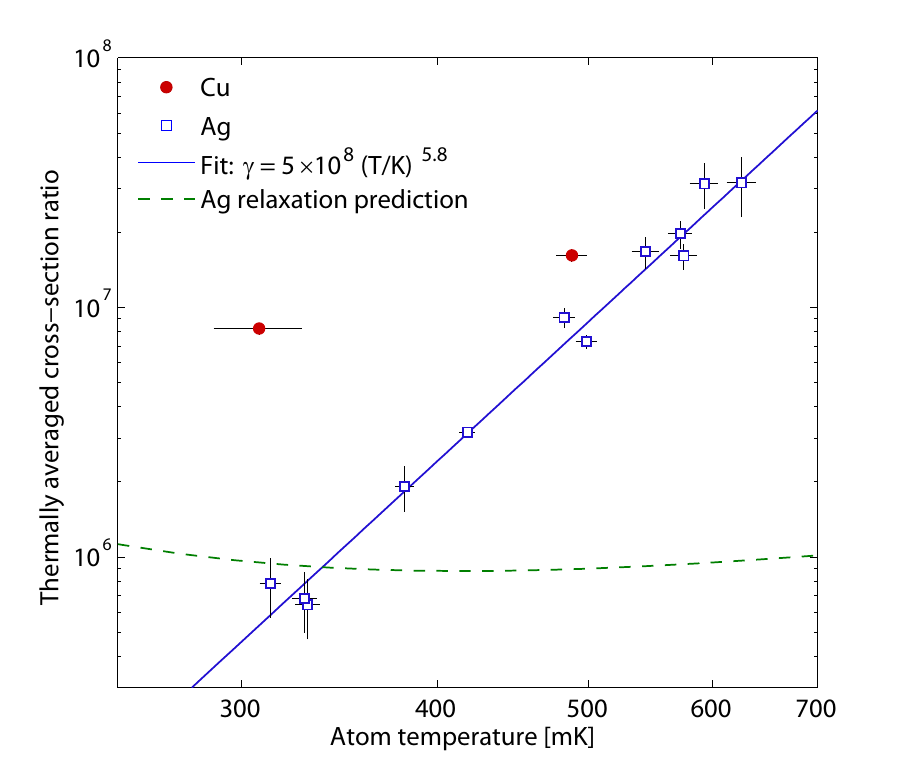}
\caption{The measured ratio of thermally averaged transport cross-section, $\bar\sigma_D$, to thermally averaged relaxation cross-section, $\bar\sigma_R$, in a 4 T trap for Cu-$^3$He and Ag-$^3$He.  The dashed line depicts a theoretical prediction for Ag-$^3$He.  Both axes are logarithmic.}
\label{gammaVsTempFigure}
\end{figure}

We investigated the dependence of $\gamma$ on temperature in the range \mbox{320 mK} to \mbox{600 mK} for Cu-$^3$He and Ag-$^3$He.  The measurement is shown in Fig. \ref{gammaVsTempFigure}.  We find a strong temperature dependence for Ag-$^3$He, with power law exponent $=5.8\pm 0.3\;\text{(stat.)}\pm 0.3\;\text{(sys.)}$.  The systematic uncertainty results from temperature measurement.  The temperature dependence of $\bar\sigma_D(T)$ for Ag-$^3$He was studied independently by measuring $\tau_0$ vs. temperature, for a constant background gas density.  We found that $\tau_0$ did not depend significantly on temperature (varying by 20\% between \mbox{420 mK} and \mbox{600 mK}).  Because the thermal dependence of $\tau_0$ is weak, we believe the strong temperature dependence of $\gamma(T)$ is due to temperature dependence of the relaxation cross-section $\bar\sigma_R(T)$.


A comparison of these data to a standard theoretical treatment {\bf shows clear disagreement between our observations and theory}.  The dominant mechanism used to explain spin-relaxation in hydrogenlike (single valence $s$ electron) atoms is the electron spin -- molecular rotation interaction \cite{hermanRubidiumRelaxation,walkerSpinRotation,walkerSpinRotation2}.  We generated a theoretical prediction for the Ag-$^3$He spin-relaxation cross-section ratio using the results of Walker {\it et al.} \cite{walkerSpinRotation} and the Ag-He potentials of Takami and Jakubek \cite{takamiAgHePotential}.  For completeness we included the contact hyperfine interaction between the electron spin and the Helium nucleus \cite{fermiHyperfine}.  The result of this calculation \cite{brahmsThesis} is plotted on the dashed line in Fig. \ref{gammaVsTempFigure}.  At low temperatures the cross section decreases as $T^{-1}$.  The weakly increasing behavior at higher temperatures is due to a scattering resonance at \mbox{0.07 meV}.

A more exact {\it ab initio} calculation of the spin-rotation and hyperfine interactions is currently underway by Tscherbul and his colleagues \cite{tscherbulPrivate}.
However, it is important to note that the power-law dependence of $\gamma$ for these interactions is expected to be less than {\bf $T^{2}$}.  This is because the strengths of these spin interaction effects are monotonically increasing functions of collision energy, save for the effect of possible scattering resonances.  Resonances would give rise to the largest possible temperature dependence, $T^{2}$.  We therefore conclude that the $\sim T^6$ dependence observed for Ag must be due to a heretofore neglected effect.

We did not observe an anomalous temperature behavior for the cross-section ratio of Cu-$^3$He.  The two data taken are consistent with the {\bf $T^{2}$} temperature dependence mentioned above.  We did not make a theoretical prediction for the Cu-$^3$He cross-section ratio because we could not find suitable internuclear potentials in the literature.

We also investigated the cross-section ratio for Ag-$^3$He vs. magnetic field.  At a constant temperature, the average field experienced by the trapped atoms scales linearly with trapping field.  At 320 mK, for $2\;\text{T}<B_\text{trap}<4\;\text{T}$ (equivalently, $ 1\;\text{T} < \langle B\rangle < 2\;\text{T}$), the cross-section ratio was measured to be $4.8\pm 0.4\times 10^6 \left ( {B_\text{trap}}/{\text{T}} \right )^{-0.9\pm 0.2}$.


We have trapped $4\times 10^{13}$ Ag atoms for up to $2.3$ s, and $3\times 10^{12}$ Cu atoms for up to $5$ s.  We developed a technique to measure transport to relaxation cross-section ratios that is independent of one's knowledge of buffer-gas density.  This ratio was measured for the Ag-$^3$He and Cu-$^3$He systems.  It was found to be large ($>10^6$) in both cases.  For Ag-$^3$He, an anomalously strong $T^{5.8}$ temperature dependence was discovered.  We have shown that this dependence is inconsistent with a standard treatment of the spin-rotation interaction.

High values of $\gamma$ indicate that collisions between these atoms and He are almost always elastic.
Historically, species observed to have high $\gamma$ have been found to have intra-atomic collisions favorable for evaporative cooling \cite{doyleEuropium,doyleChromium}. 
The elasticity of intra-species collisions could be observed by removing the buffer gas.  In our experiment, the trap $\eta$ of 4 to 8 was too low to allow for removal of the buffer gas \footnote{
In buffer gas cooling experiments, an $\eta \geq 10$ is required to remove the buffer gas.  Cf. \cite{brahmsThesis}.
}.
By decreasing $T$ to 270 mK or by using a slightly deeper magnetic trap,
the buffer gas could be removed and evaporative cooling could be attempted.  If this is successful, dense samples of ultracold copper or silver could be produced using buffer gas loading.


We are very grateful to Roman Krems, Kate Kirby, Hossein Sadeghpour, Alex Dalgarno, and Jim Babb for many helpful discussions in understanding spin-relaxation processes and their theoretical calculation.
This work was made possible by grants from the National Science Foundation and the Office of Naval Research.

\bibliographystyle{apsrev}
\bibliography{agRelaxation}

\end{document}